\documentclass{aa}
\usepackage{graphicx}
\begin{document}
   \title{Photo-Luminescence and Possible Forsterite nanoparticles\\
          Model of Extended Red Emission
}


\titlerunning{Photo-Luminescence and Forsterite\\ nanoparticle Model
               of ERE} 

   \author{K. Koike\inst{1}
          \and
          H. Chihara\inst{2,4}
          \and
          C. Koike\inst{2}
          \and
          M. Nakagawa\inst{1}
          \and
          M. Okada\inst{3}
          \and\\
          M. Matsumura\inst{1}
          \and
          J.Takada\inst{3} 
}

   \offprints{K. Koike}

   \institute{Faculty of Education, Kagawa University,
             Takamatsu 760-8522, Japan\\
             \email{Koike@ed.Kagawa-u.ac.jp}
         \and
             Kyoto Pharmaceutical University,
             Kyoto 607-8412, Japan\
         \and 
             Research Reactor Institute, Kyoto University,
             Kumatori 590-0499, Japan\
         \and
             Department of Earth and Space Science, 
             Osaka University, 
             Toyonaka, Osaka 560-0043, Japan\
}


\abstract{
 Possible forsterite nanoparticle model of the Extended Red Emission(ERE) is 
proposed on the basis of photo-luminescence of forsterite after gamma-ray and 
neutron irradiation. 
 Forsterite exhibits interesting thermoluminescence spectrum similar to ERE of Red
Rectangle after irradiation in low temperature. It is shown that the forsterite 
after thermoluminescence is over exhibits photo-luminescence(PL) when Ultraviolet 
ray is irradiated. The structure of PL spectrum is almost similar to that of 
thermoluminescence. In order to explain
small variations of the peak position of wavelength of ERE spectrum, possible 
nanoparticle model of forsterite is investigated. Our model is consistent to 
the ISO observation data in near and middle infrared region, which suggest the 
existence of forsterite. 
   }
\maketitle
%

\section{Introduction}

 Extended red emission (ERE) is a broad emission band with a peak wavelength 
between 600 and 850 nm, and with a width between 60 and 120 nm seen in many 
dusty astrophysical objects such as reflection nebulae, planetary nebulae, 
HII regions, halos of galaxies, and even in the Diffuse Interstellar Medium. 
The observation of ERE in the Diffuse Interstellar Medium shows that ERE 
is a general phenomenon. 
Though the carrier for ERE is not yet clear,
the proposed carriers are
hydrogenate amorphous carbon (HAC), quenched carbonaceous 
composite (QCC), $\rm C_{60}$, carbon nanoparticles, polycyclic aromatic 
hydrocarbons (PAHs),  and silicon nanoparticles, and most of them appear to 
be unable to explain the observed ERE spectra (see Ledoux et al. 
2001; Witt et al. 1998, for a summary).

We have suggested that thermoluminescence spectra of forsterite 
after $\gamma$-ray irradiation are very similar to ERE of Red Rectangle
 (Koike K. et al., 2002, here after KK), 
and have discussed such possibility that thermoluminescence 
is related to the changes of property of interstellar and circumstellar 
matter by various irradiation in that space.
It is, however, not so plausible that the irradiation energy is sufficient
to explain such emissions.

 Recently, we have found that the forsterite after thermoluminescence 
is over exhibits photo-luminescence(PL) when ultraviolet ray (UV)
is irradiated. This fact seems to suggest a possible realistic mechanism 
of ERE.
This paper is concerning to this problem.

   \begin{figure*}
   \includegraphics[width=17cm,height=8cm]{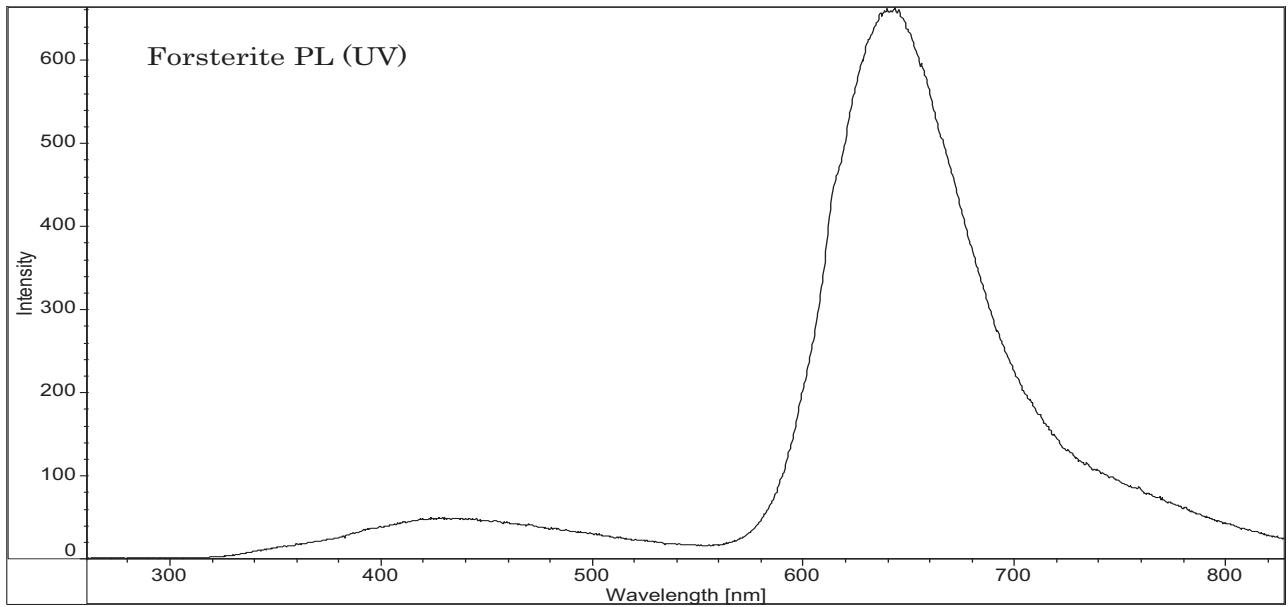}
   \caption{Photoluminescence spectra of forsterite $(\rm Mg_2SiO_4)$ single 
         crystals under UV irradiation. The sample is firstly irradiated
         by $(\gamma)$-ray in liquid nitrogen, and warmed till about 500K.
         After thermoluminescence has been over, it is irradiated by
         UV of a mercury lamp at room temperature.}
\label{Forste_UV_R}
    \end{figure*}

Interstellar and circumstellar matter is irradiated by high energy 
electromagnetic and 
cosmic ray particles such as $\gamma$ rays, neutrons, 
protons and heavy-ions etc. These irradiation will cause some changes 
on properties such as optical ones of these materials. 
Especially, it is known that extremely large fluxes of neutrons and 
$\gamma$-rays have been emitted during super-nova explosions.
Furthermore, interstellar and circumstellar space is typically at 
extremely low temperature and is always irradiated 
by electromagnetic radiation and by cosmic ray particles
for long time-scale.
The effect of this radiation will accumulate in the low temperature environment.
In the circumstellar region of both young and evolved stars as well as 
in the solar system, forsterite and enstatite have been found (Waters et 
al., 1998ab ; Malfait et al., 1998; Wooden et al., 1999) by the Infrared 
Space Observatory (ISO) (Kessler et al., 1996). 
Carbonates such as dolomite 
$\rm CaMg(CO_{3})_{2}$, breunnerite $\rm Mg(Fe,Mn)(CO_{3})_{2}$, 
calcite $\rm CaCO_3$, and $\rm Mg$,$\rm Ca$-bearing siderite
$\rm FeCO_3$ were found in CI 
chondrite (Endress, Zinner and Bischoff, 1996). 
Among these carbonates, Ca-bearing minerals such as dolomite and calcite were 
also detected in dust shells around evolved stars by ISO (Kemper et al., 2002). 
Especially, it should be noted that the broad emission feature responsible for
extended red emission (ERE) 
appears at about the 600 -- 900 nm region in many reflection nebulae,
and among reflection nebulae the Red Rectangle nebula shows stronger 
intensity by one order(Witt and Boroson, 1990). In the Red Rectangle nebula,
both PAH- and crystalline silicates (forsterite and enstatite)-features 
were observed by ISO (Waters et al., 1998b).
We have suggested, on the basis of the measurement of thermoluminescence 
for irradiated silicates and carbonates, our thermoluminescence spectra 
of forsterite at 645--655 nm is very similar to the ERE of the 
Red Rectangle (KK, 2002).

In this paper, we will investigate the changes of properties of 
silicates by irradiation in the context of astrophysics. Especially, it is 
emphasized that the forsterite after thermoluminescence 
is over exhibits photo-luminescence(PL) when ultraviolet ray (UV)
is irradiated. The structure of PL spectrum is almost similar to that of 
thermoluminescence.
Further, possible forsterite nanoparticle model 
of ERE is also discussed on the basis of this fact.

  \begin{figure}
  \centering
   \includegraphics[width=7.5cm]{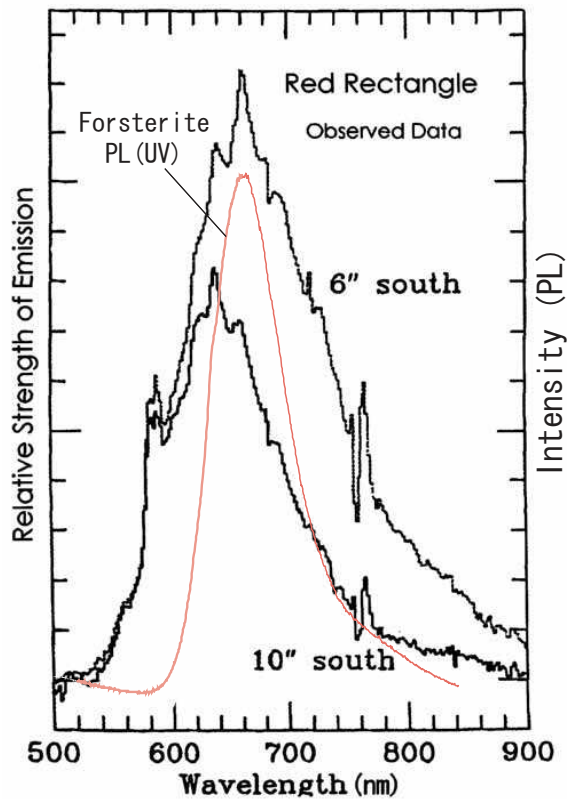}
 \caption{Comparison with Observation Data of Red Rectangle Spectrum
  (From Witt A.N. and Boroson T.A. 1990) and Photoluminescence spectra of 
   forsterite under UV irradiation (Fig.1) }
 \label{Forste_UV-Red}
 \end{figure}

\section{Photoluminescence of forsterite under UV irradiation}

 We have ever discussed that thermoluminescence spectrum of irradiated 
bulk samples of forsterite, $\rm Mg_2SiO_4$, is very similar to the ERE of
Red Rectangle.
The samples were irradiated with gamma-rays to a dose of about 
$10.4 \times 10^4$ Gy $\rm(J/Kg)$ in liquid nitrogen 
using the $\rm ^{60}Co$ gamma-ray irradiation facility of Kyoto University 
Reactor. The gamma-rays of $\rm ^{60}Co$ have two peaks at 1.1 MeV and 
1.3 MeV. 
Our samples of forsterite were synthesized by Takei and Kobayashi (1974), and 
Tachibana (2000) using the CZ (Czochralski) method, with high accuracy. 
The bulk of the irradiated forsterite is  about 138 mg. 
We have measured the thermoluminescence 
spectra of these samples using a spectrophotometer
(including a CCD camera, from Princeton Instruments, Inc.). 
The sample is put on a thermally-isolated plate, 
previously cooled to liquid nitrogen temperature. 
The luminescence emission during warming is introduced to the CCD measuring 
system using an optically transparent fiber.
We measured the thermoluminescence spectrum of forsterite from liquid nitrogen
temperature to about 500K. As is discussed previously, the spectrum is very 
similar to ERE of Red Rectangle.
We have discussed such mechanism that the effect of various irradiation
on cosmic matter is accumulated in the extremely low temperature environment of
interstellar and circumstellar space,
and it will only be observed provided that the condition to release the
accumulated energy is realized in circumstellar space. This may occur when 
irradiated dust moves to a warmer domain in an interstellar or circumstellar 
environment.

However, amount of irradiation energy seems to be insufficient to explain the 
large emission energy of ERE.
This difficulty will be overcome by taking into account of the interesting fact 
that we have recently found. That is, the forsterite after thermoluminescence 
is over exhibits photo-luminescence(PL) when ultraviolet-ray (UV) is irradiated.
For the UV-ray source, we have used  a mercury lamp.

 Fig.\ref{Forste_UV_R} shows the photo-luminescence spectra 
of forsterite $(\rm Mg_2SiO_4)$ under UV irradiation at room temperature
after the thermoluminescence of $\gamma$-ray irradiation is over.
It should be noted that this spectra is almost similar to that of
thermoluminescence, however, the detailed structures of spectrum seems to 
disappear.
Does photo-luminescence of forsterite occur when only UV-ray is irradiated
without the preceding $gamma$-ray irradiation on the sample?
The answer is no!
We have confirmed that any photo-luminescence does not appear 
for $\gamma$-ray non-irradiated forsterite.

Furthermore, at liquid nitrogen temperature we have investigated the effect of 
UV-ray irradiation on forsterite after thermoluminescence caused by 
$\gamma$-ray irradiation had been over. In this temperature, we
cannot find UV photo-luminescence. However, when the UV irradiated forsterite is 
warmed, thermoluminescence spectra similar to the $\gamma$-ray irradiated one 
appears.
The UV photo-luminescence spectrum at fixed temperature between
the liquid nitrogen and room temperatures are not yet measured, because it is 
very difficult to hold the sample at the fixed temperature in this range. 

It should be emphasized that the existence of  UV emission from almost 
all stars or nebulae is well known fact, and our discovery of UV photo-luminescence 
of forsterite seems to resolve the problem of energy source. It should be noted, 
however, that peaks of spectrum of various ERE exhibit small amount of difference. 
It is just a nanoparticle model of ERE that can explain  shift of the peak position 
of spectrum.

\section{Possible silicon nanoparticle model and some problems}

It is known that a prominent model of ERE is the silicon nanoparticle model.
 The general mechanism of PL in solids is summarized as follows. When a photon of 
energy higher than the so-called band gap of the material (the separation between 
valence and conduction bands) is absorbed, an electron-hole pair is created. On a 
very short timescale the energy is thermalized so that the energy separation 
between the electron and the hole becomes approximately equal to the energy of 
the gap. Then the pair can radioactively recombine, giving rise to the 
photo-luminescence. As a result, the peak position of the photo-luminescence 
roughly reflects the band gap of the material. 
It should be noted that for the energy gap $V_{g}$ [eV], the corresponding 
wavelength $\lambda$ [nm] is given as,

\begin{center}
\begin{displaymath}
~~~~~~~~~~~~~~~~~~~~~~~~~\lambda = 1234~/~{V_{g}} 
\label{eq:Wavelength}
\end{displaymath}
\end{center}
in the degree of $10^{-4}$, as is easily confirmed from $h\nu = E$. Then, the 
energy gap corresponding to ERE wavelength is about 1.4 - 2.2 eV.

In metals there is no PL because there is no gap. For semiconductor bulk silicon, 
the energy gap is about 1.17 eV.  In nano sized silicon particles,  the transition 
probability is drastically enhanced and the energy gap is extended to ERE range
because the surfaces of nanoparticles is completely passivated to avoid quenching 
the photo-luminescence and the spatial confinement is associated with a broadening 
of the pair state in momentum space.

The most important consequence of the spatial confinement of the electronic wave 
function in nano sized systems is the progressive widening of the band gap as the 
particle size is reduced. Theoretical studies have shown (Delerue et al. 1993) in 
quantitative this remarkable fact in a certain kind of model. 
The physical meaning of this fact seems
to understand by making use of uncertainty principle of quantum system 
qualitatively.
That is, the spatial confinement of the electronic wave function restricts the 
uncertainty of electron $\Delta X$ to the diameter of restricted size of the 
nanoparticle, then due to the uncertainty principle,
\begin{center}
\begin{displaymath}
~~~~~~~~~~~~~~~~~~~~~~~~~\Delta X \Delta P \ge  \hbar 
\label{eq:Uncertainty}
\end{displaymath}
\end{center}

\noindent
the magnitude of $\Delta P$ becomes large. Thus the smaller the size of particles, 
the larger the energy scale is extended.
Does the silicon nanoparticle exist really? It is not probable that all silicons 
can exist only in nanoparticle state in universe. That is, it is natural to
suppose that the conditions to form the more large particles or bulk of silicon
will be often realized  in universe where embrace various conditions.
As discussed before, the energy gap is extended in the nano sized silicon 
particles. In the more large silicon particles or small bulk, it is expected,
from simulation experiments, that the strong emission feature will appear at
16.4 $\mu m$ and the peak of silicon spectrum at 20 $\mu m$ in the near and middle
infrared-ray region. However, both of them are not yet observed until now. 
In practice, ISO result show the characteristic four peaks which exhibits the 
existence of forsterite.

\section{Investigation of formation process of Elements, and Silicates}

In order to investigate possible existence form of silicates,
let us examine the formation process of elements in stars. 
In the core of proto-star, firstly  H burns through the thermo-nuclear fusion. 
This process proceeds very slowly and He is formated. At a certain temperature,
He-burn starts and C and O are produced in the core of star. According to 
rising the temperature of core, C-burn starts and it produces Mg etc in its core 
region. Further, O-burn starts and produces Si in the core of star.
In a star with heavy mass, Si burns finally in core and produce Fe etc.
 It should be noted that Si is produced in the core region, which is already 
surrounded by C and O in its shell structure. Further, in a certain kind of heavy 
stars, it is known that heavy elements than Fe is produced through the r- and 
s-processes.

By thermonuclear runaway called as thermal pulses, the entire envelop is mixed 
efficiently and Si is transported to the surface of the star. Important aspect 
for stars on the AGB is that these stars lose a large fraction of their mass. 
The physical mechanism responsible for these high mass-loss is thought to be 
stellar pulsation in combination with the formation of dust. Further, the stellar
wind reduces the envelop mass of star. 

From the above summarization of formation condition of dust based on nuclear fusion
in star, it seems to be natural to suppose that almost all Si exists in the form of
silicates such as  $\rm SiO_2$, $\rm MgSiO_3$, $\rm Mg_2SiO_4$, or  $\rm SiC$,
so far as there is no special mechanism which prevents to make these crystalline.

\section{Possible forsterite nanoparticle hypothesis}

Why the $\gamma$-ray or neutron irradiated forsterite exhibits the 
photo-luminescence under UV irradiation? Though the change of material property
of forsterite by irradiation is not yet clear except that of optical
property, it seems that the new energy band-like structure is formed by lattice
defects and/or deformations of the structure as irradiation effects. 
It should be noted that the new energy level or band-like structure is formed
by lattice defect in some insulators such as $\rm TiO_2$, and they have properties 
of semiconductor.
The mechanism of UV photo-luminescence of forsterite seems to be realized through 
the similar mechanism.
In order to make clear irradiation effect on forsterite by the method of
solid state physics, it is expected to investigate to the change of forsterite 
structure by both ESR experiments and theoretical approach.

Summarizing our investigation,  we will propose possible forsterite model of 
ERE, in which ERE is caused by UV photo-luminescence of various size of forsterite
containing nano size particles, where these forsterite particles have been 
irradiated by various 
kind of cosmic-rays. The variety of peak of emission spectrum is due to the 
existence of nano size particle of forsterite.
In fact, the existence of forsterite is confirmed by ISO spectrum.

\section{Discussion}

 It should be noted that interstellar and circumstellar space is typically at 
extremely low temperature and is always irradiated by both electromagnetic 
radiation and by cosmic ray particles over cosmological timescales. 
Furthermore, it is well known that extremely large fluxes of neutrons and 
gamma-rays are emitted during supernova explosions.
 We have ever suggested that thermoluminescence spectra 
of forsterite after $\gamma$-ray irradiation are very similar to ERE of Red 
Rectangle. However, the energy efficiency of thermoluminescence seems to be not
sufficient to explain ERE. In this paper, we have discussed on the fact
that the photo-luminescence of irradiated forsterite exhibits similar spectrum
to that of thermoluminescence.
It should also be remembered that forsterite and enstatite have been found by many 
ISO observations in many oxygen-rich young and evolved stars.

 Is the irradiated forsterite possible to be really one of the carrier of ERE?
It is possible so far as its temperature is below about 1000K, because
it is known in many insulators the effects of irradiation are almost maintained.
However, the irradiation effects on insulators are known to fade out by annealing 
it to high temperature above this degree. In such case, can forsterite be a carrier 
of ERE?

In semiconductor silicon, it is well known that the existence of a small amount
of impurity caused by a certain kind of minor elements such as As, P, B, Ga or In 
etc cause new energy levels and forms new type of semiconductors. Though little 
is known about forsterite, it seems to be possible that a certain kind of impurities
cause some semiconductor-like structure of forsterite. Investigation of such 
possibility is further problem.
In this context, it should be noted that the melting point of silicon
is $1410^{\circ}C$, while that of forsterite is very high, and is $1890^{\circ}C$.
It is known that the existence of forsterite is observed in addition to ISO 
spectrum, in meteorites, while silicon itself is not found in both of them. 
Thus, together with the investigation of formation condition of silicates in the 
previous section, the silicon nanoparticle model seems to be highly hypothetical 
one in the present stage.

We can get easily high-quality crystalline silicon because it is the most 
fundamental semiconductor, where the solid state properties is very well studied 
together with the problem of impurity. 
As for the forsterite, it is very difficult to get the high-quality one.
Then, we have used the synthesized one by CZ method in laboratory. 
For the impurity component in our forsterite sample, Takei and Kobayashi
(\cite{takei}) reported  previously that spectrographic analysis shows that the 
forsterite sample is pure. 
However, neutron activation analysis reveals that it contains an infinitesimal 
quantity of Ir at about 16--18 wt ppm. This level of Ir impurity is below the 
detection limit of spectrographic analysis.
We have also measured our sample using radio activation analysis, 
and confirmed that our 
sample is almost pure; that is, other elements except for $\rm Mg, Si, O$ 
and an infinitesimal quantity of Ir are not detected.

In order to confirm the reappearance of thermoluminescence of forsterite in 
other sample, we have get forsterite powder for china and porcelain from a pottery 
"Marusu" (Japan),
where details of the method of creation of it is not open.
We have investigated thermoluminescence of the "Marusu" forsterite, and get
almost the same result as our synthesized one.

\section{Summary}

In this paper, we have discussed about the possibility that the photo-luminescence
of irradiated forsterite by $\gamma$-ray and cosmic-rays may contribute to ERE.
It should also be re-emphasized that forsterite and enstatite have been found 
by many ISO observations in many oxygen-rich young and evolved stars.
It is natural to suppose they assume any responsibility in many phenomena such as 
ERE. In order to investigate further properties of forsterite, much lot of 
crystalline forsterite with high-quality is necessary, just as crystalline silicon 
which is mass-produced as the important material of semiconductor. 
Synthesizing forsterite
and enstatite in laboratory is very restricted in practice. It is expected to 
investigate properties of forsterite with a certain kind of minor elements as 
impurity. 

It is interesting to suppose the case in which silicon nanoparticles and irradiated
forsterite co-exist as the carrier of ERE. In such case,  almost all silicon in the
circumstellar space should exist as nanoparticle form, while forsterite is possible
in various size from bulk to nanoparticle, as is suggested from the restriction 
of present observation such as ISO etc. 

Finally, investigating the detailed structure of change on forsterite by irradiation
and possible semiconductor-like structure caused by impurity in the context of 
astrophysics is further problem.


\begin{acknowledgements}
  
 This work was supported by the KUR projects (14P5-6,~~15P6-6).
 Part of this work was supported by Grant-in-Aid of Japanese Ministry of 
 Education, Science, and Culture (15540384 ,~~12440054). 
\end{acknowledgements}


\begin{thebibliography}{}

\bibitem[1996]{endress} Endress M., Zinner E. \& Bischoff A. 1996,
        Nature, 379, 701

\bibitem[1993]{Delerue} C. Delerue, G. Allan and M.Lannoo 1993,
        Physical Review B, 48, 11024

\bibitem[2002]{kemper}KEMPER F., Jaeger C., Waters L.B.F.M., et al., 2002, 
       Nature, 415, 295

\bibitem[1996]{kessler}KESSLER M.F., STEINZ J.A., ANDEREGG M.E., et al.1996,
       A\&A, 315, L27

\bibitem[2000-1]{koikeC1} Koike C., Tsuchiyama A., Shibai H. et al. 2000-1,
        A\&A, 363, 1115

\bibitem[2000-2]{koikeC2} Koike C., Chihara H., Tsuchiyama A. et al. 2000-2,
        Proc. 33rd ISAS Lunar and Planet. Symp. 33, 95

\bibitem[2002]{koikeC} Koike C., Chihara H., Koike K. et al. 2002,
        M\&PS, 37, 1151

\bibitem[2002]{KoikeK} Koike K., Nakagawa M., Koike C., Okada M. 
        and Chihara H. 2002,
        A\&A, 390, 1133

\bibitem[2001]{Ledoux} Ledoux G., Guillois O., Huisken F. et al. 2001,
        A\&A, 377, 707
        
\bibitem[1998]{malfait} Malfait K., Waelkens C., Waters L.B.F.M. et al. 1998,
        A\&A, 332, L25
        
\bibitem[2000]{tachibana} Tachibana S., 2000, Ph. D. Thesis, Osaka 
       University

\bibitem[1974]{takei} Takei H. and Kobayashi T. 1974,
      J. Crystal Growth, 23, 121
      
\bibitem[1998a]{waters} Waters L.B.F.M., Beintem D.A., 
      Zijlstra A.A. et al. 1998,
      A\&A, 331, L61

\bibitem[1998b]{waters2} Waters L.B.F.M., Waelkens C., Van Winckel H. et al.,
     1998, Nature, 391, 868

\bibitem[1990]{witt} Witt A.N. and Boroson T.A. 1990,
        ApJ, 355, 182

\bibitem[1998]{witt2} Witt A.N. , Gordon K.D. and Furton D.G. 1998, 
        ApJ, 501, L111.

\bibitem[1999]{wooden} Wooden D.H., Harker D.E., Woodward C.E. et al. 1999,
        ApJ, 517, 1034

\end{thebibliography}
\end{document}